# Seebeck Effect in Magnetic Tunnel Junctions

Marvin Walter, Jakob Walowski, Vladyslav Zbarsky, Markus Münzenberg*
*I. Physikalisches Institut, Georg-August-Universität Göttingen, Germany*

Markus Schäfers, Daniel Ebke, Günter Reiss,
*Department of Physics, Universität Bielefeld, Germany*

Andy Thomas
*Department of Physics, Universität Bielefeld, Germany and Institut für angewandte Physik, Hamburg, Germany*

Patrick Peretzki, Michael Seibt,
*IV. Physikalisches Institut, Georg-August-Universität Göttingen, Germany*

Jagadeesh S. Moodera, *Massachusetts Institute of Technology (MIT), Cambridge, USA*

Michael Czerner, Michael Bachmann, Christian Heiliger, *I. Physikalisches Institut, Justus-Liebig-Universität Gießen, Germany*

## Abstract

Creating temperature gradients in magnetic nanostructures has resulted in a new research direction, i.e., the combination of magneto- and thermoelectric effects [1,2,3,4,5]. Here, we demonstrate the observation of one important effect of this class: the magneto-Seebeck effect. It is observed when a magnetic configuration changes the charge based Seebeck coefficient. In particular, the Seebeck coefficient changes during the transition from a parallel to an antiparallel magnetic configuration in a tunnel junction. In that respect, it is the analog to the tunneling magnetoresistance. The Seebeck coefficients in parallel and antiparallel configuration are in the order of the voltages known from the charge-Seebeck effect. The size and sign of the effect can be controlled by the composition of the electrodes' atomic layers adjacent to the barrier and the temperature. The geometric center of the electronic density of states relative to the Fermi level determines the size of the Seebeck effect. Experimentally, we realized 8.8 % magneto-Seebeck effect, which results from a voltage change of about -8.7 $\mu$V/K from the antiparallel to the parallel direction close to the predicted value of -12.1 $\mu$V/K. In contrast to the spin-Seebeck effect it can be measured as a voltage change directly without conversion of a spin current.

*Corresponding author: Markus Münzenberg, e-mail: mmuenze@gwdg.de

The creation of an electric field by a temperature gradient in a material is known as Seebeck effect since 1826. In the last years new spin dependent thermal effects have been discovered in ferromagnets and the Seebeck effect receives novel interest. Gravier et. al. described the transport of heat and spin in magnetic nanostructures [1]. Uchida et. al. experimentally found the spin-Seebeck effect [2] driving this novel field, e.g., in nanoscale metal structures [3], in magnetic insulators and semiconductors [4,5]. A strong asymmetry of the density of states with respect to the Fermi level promotes the heat-driven electron transport that leads to the common charge-Seebeck effect. These strong asymmetries can be found in the spin split density of states in ferromagnetic materials. Previously, the effect amplitude resulting from this spin asymmetry was believed to be a second-order effect [6]. In this work, we demonstrate that the magneto-Seebeck effect can be large. We first present *ab initio* calculations which show that this effect can be of the same order as the charge-Seebeck effect. Using magnetic tunnel junctions, where two ferromagnets are separated by a thin insulating tunnel barrier. The effect is related to a half metallic behavior of the tunnel junction transmission. Our experiments show that a thermoelectric power can be generated in such nanostructures over distances of only 2.1 nm, the thickness of the tunnel barrier. The change from parallel to antiparallel electrode configuration is -8.7 µV/K at room temperature, while maintaining all other conditions in the junction constant. Related with this magnetization switching we calculated a magneto-Seebeck effect of 8.8 %. In theory, this change is predicted to be up to 1000 % / 100 µV/K [7]. In future spincaloritronic [8] applications, the local cooling of an individual nanometer-sized area could, therefore, be switched magnetically. The junction size allows stacking and nano-integration of these novel thermopower devices.

The magneto-thermal effect is based on the seminal work conducted by Johnson and Silsbee [9]. They gave a general description of the mechanisms that affect a ferromagnetic material when a heat flow causes a temperature gradient. Strong thermomagnetic effects can be expected in a half metal where the spin polarization can be up to 100% [10]. One can define a spin-dependent Seebeck coefficient by replacing the charge-dependent Seebeck voltage by a voltage generated for each spin channel. The difference between the two spin-dependent Seebeck coefficients is driving a spin accumulation. In contrast, the magneto-Seebeck effect is different from the spin-Seebeck effect, because it is not related to a spin-voltage generation. It occurs in junctions and is similar to the giant (GMR) and tunneling magnetoresistance (TMR). It results in a charge-Seebeck effect that is changed by the magnetic orientation of the electrodes. This voltage is accessible directly without conversion. In order to have a high charge-Seebeck effect, a high asymmetry in the energy dependence with respect to the electrochemical potential for the transport states is necessary, realized in semiconductors as shown in Fig.1 a). Consequently, for the thermomagnetic effect, these energy asymmetries must be different for spin up and spin down carriers. For our experiments, the recent progress in giant tunneling magnetoresistive (giant TMR) junctions enabled us to use magnetic tunnel junctions (MTJs) with high spin asymmetry. Their large contrast in the spin-dependent transmission due to different symmetries of the tunneling states in both spin channels should lead also to different energy asymmetries of the tunneling states as shown in Fig. 1 b). We define the magneto-Seebeck ratio ($S_{MS}$) from the Seebeck coefficients in the parallel and anti-parallel configuration, as given in equation (1):

$$S_{\text{MS}} = \frac{S_{\text{P}} - S_{\text{AP}}}{\min(S_{\text{P}}, S_{\text{AP}})} \qquad (1)$$

At first glance, it seems that the magnetoresistance, the spin-Seebeck effect, and the magneto-Seebeck effect should be related to each other. However, these are different effects, and, in general, it is not possible to calculate one from the others.

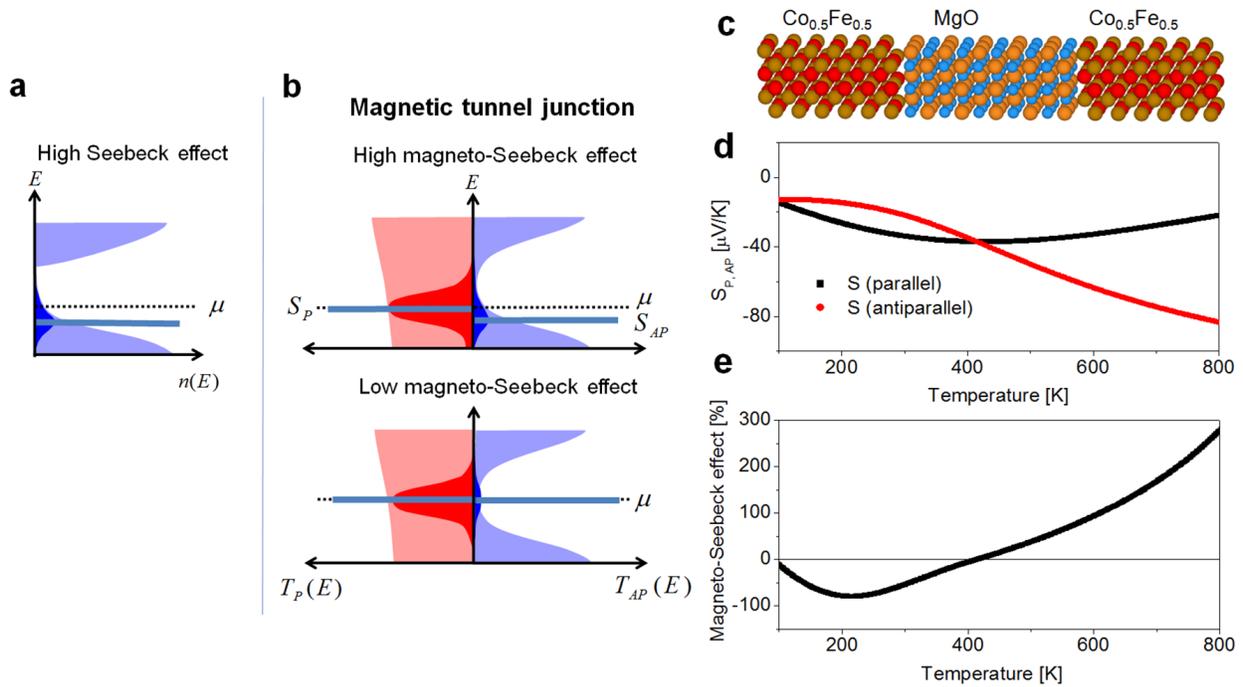

**Figure 1 Origin of the magneto-Seebeck effect.** a) Semiconductors are known to generate high Seebeck effects. b) In magnetic tunnel junctions, thermal differences in the electron distributions and strong asymmetry in the spin-dependent tunneling channels are depicted. T(E) is the transmission of the full tunnel junction for which either the ferromagnetic electrodes can be a highly spin polarized half metal or the combination of the barrier and the ferromagnet exhibit half-metallic characteristics. The function $T(E)(-\partial_E f(E,\mu,T))$ is given in darker color. The thick line marks the resulting value of the geometric center $S_P$ and $S_{AP}$. In the lower symmetric case, the magneto-Seebeck effect is vanishing. c) Calculation of the Seebeck coefficients as a function of temperature for tunnel junctions with 10 monolayers of MgO as a barrier. The magnetic layers are 20 monolayers thick. The semi-infinite leads are Cu in the bcc-Fe structure. We assume a mixed termination of FeCo at the FeCo/MgO interface that is an ordered, 2x1, in-plane supercell with one Fe and one Co atom. d) Seebeck coefficients for the parallel (P) configuration and the antiparallel (AP) configuration are shown, e) is the corresponding magneto-Seebeck effect $S_{\text{MS}}$.

To understand this point, it is important to realize that the transport coefficients are calculated from the transmission function T(E) of the tunnel junction but that they have different integral values. The conductance g is determined by the integral of the transmission function T(E) multiplied by the derivative of the electron occupation function $\partial_E f(E,\mu,T)$ at temperature T and electrochemical potential μ:

$$g = \frac{e^2}{h} \int T(E)(-\partial_E f(E,\mu,T))dE \qquad (2)$$

The Seebeck coefficient is also given by the transmission function T(E) multiplied by the derivative of the occupation function $\partial_E f(E,\mu,T)$:

$$S = -\frac{\int T(E)(E-\mu)(-\partial_E f(E,\mu,T))dE}{eT \int T(E)(-\partial_E f(E,\mu,T))dE} \qquad (3)$$

Different to the magnetoresistance, the Seebeck coefficient is the geometric center of $T(E)(-\partial_E f(E,\mu,T))$. Fig. 1 b) illustrates these quantities for two different cases. The geometric center for parallel and antiparallel configuration ($S_P$ and $S_{AP}$) is marked by the thick line. We assume a transmission function that has different energy asymmetries in both magnetic configurations and different positions of the electrochemical potential. In the first case, a high TMR and a high magneto-Seebeck ratio are obtained. In the second case, the value of $S_{MS}$ is essentially zero, but the TMR is highest. Generally speaking, cases with vanishing value of $S_{MS}$ and large TMR (or vice versa) are also possible. Therefore, we can tailor magnetic tunnel junctions to be good candidates for large magneto-Seebeck effects. Consequently, we investigated temperature-induced voltages in magnetic tunnel junctions starting with samples showing large TMR ratios. Two different types of junctions with large TMR values could be used, i.e., Fe-Co/MgO/Fe-Co and half-metallic compounds. We focus on the former case, since it is demonstrated to have the largest experimental value of 604% at room temperature [11]. The tunneling states of the electrons have been thoroughly investigated for MgO-based magnetic tunnel junctions and the understanding of spin polarization of the current and the quantitative approach to magnetoresistance in tunnel junctions has advanced enormously in recent years.

Our theoretical investigations are *ab initio* calculations based on density functional theory. In particular, we used the Korringa-Kohn-Rostoker and the non-equilibrium Green's function method to obtain the transmission function T(E) [12]. Using T(E), we calculated the transport coefficients according to Eqs. (2) and (3) [13, 14]. We investigated the magneto-Seebeck coefficients for different temperatures for $Fe_{0.5}Co_{0.5}$/MgO/$Fe_{0.5}Co_{0.5}$ magnetic tunnel junctions with bcc structure of the ferromagnetic electrodes. The temperature dependence is considered only within the electron-occupation function. Due to coherent tunneling, the atomic structure of the interface could be important. Therefore, we investigated the Seebeck coefficients for different possible interface structures, i.e., the Fe-terminated structure, the Co-terminated structure, and a mixed-termination structure. The results at a temperature of 300 K listed in Table 1 show a strong dependence on the interface structure. Even a sign change was observed. However, the case that the layer next to the barrier is pure Co or pure Fe is unlikely in the experiment. Consequently, we continued our investigation with the mixed termination structure ($Co_{0.5}Fe_{0.5}$). In Fig. 1 d), $S_P$ and $S_{AP}$ are plotted as a function of temperature for a tunnel junction that has an MgO barrier that was 10 monolayers thick. In addition, we plot the corresponding magneto-Seebeck ratios (Fig. 1 e). Although $S_P$ and $S_{AP}$ do not change sign, $S_{MS}$ does when $S_P = S_{AP}$. We found that $S_P$ and $S_{AP}$ were large when compared to charge-Seebeck coefficients.

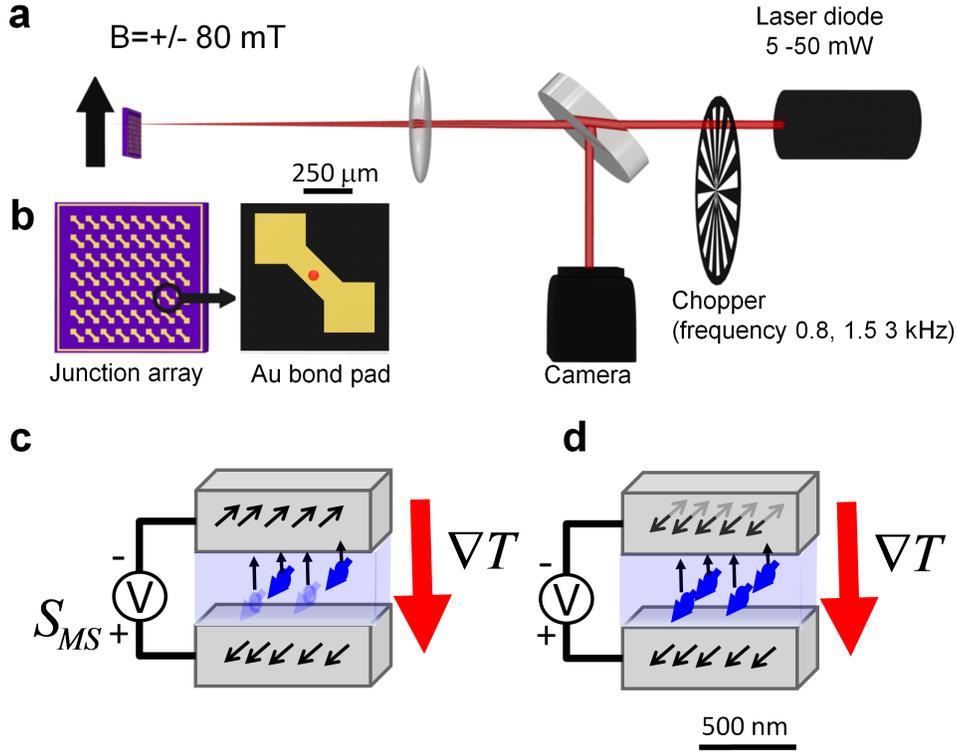

**Figure 2 Switching of the Seebeck effect via the magnetization.** a) Schematics of the laser heating setup and b) Au top contact geometry of the device with the laser spot dimensions. c) From antiparallel d) to parallel orientation of the layer magnetization, the charge-Seebeck voltage varies. By the magnitude of its change the magneto-Seebeck effect $S_{MS}$ is defined.

Table 1: The Seebeck coefficients for parallel $S_P$ and antiparallel configuration $S_{AP}$ and the magneto-Seebeck effects calculated for different supercells at a temperature of 300 K. The results show the sensitivity to the interface composition. $S_{MS}$ defines the relative change and can be negative or positive.

**FeCo/MgO/FeCo with a 10-monolayer MgO barrier**

|  | $S_P$ [µV/K] | $S_{AP}$ [µV/K] | $S_P - S_{AP}$ [µV/K] | $S_{MS}$ [%] |
|---|---|---|---|---|
| CoFe | -19.7 | -32.4 | 12.7 | 64.1 |
| FeCo | 45.9 | -50.0 | 95.9 | 209.0 |
| CFFC | 9.4 | -44.6 | 54.0 | 573.2 |
| $Co_{0.5}Fe_{0.5}$ | -34.0 | -21.9 | -12.1 | -55.2 |
| Experimental value | -107.9 (-1300) | -99.2 (-1195) | -8.7 (-105) | -8.8 (-8.8) |

Abbreviations: CoFe - $Co_{0.5}Fe_{0.5}$ layers with Co at the MgO interface. FeCo - $Co_{0.5}Fe_{0.5}$ layers with Fe at the MgO interface. CFFC - $Co_{0.5}Fe_{0.5}$ layers with Fe at one of the MgO interfaces and Co at the other. $Co_{0.5}Fe_{0.5}$ - super cell in-plane with Co:Fe 1:1 at the interface. The values derived from the experiment are given for a temperature difference at the MgO barrier of 53 mK (4.4 mK) respectively. The temperature difference ΔT is taken from the numerical simulation of the temperature gradients using the thin film value (bulk value) of the thermal conductivity of MgO.

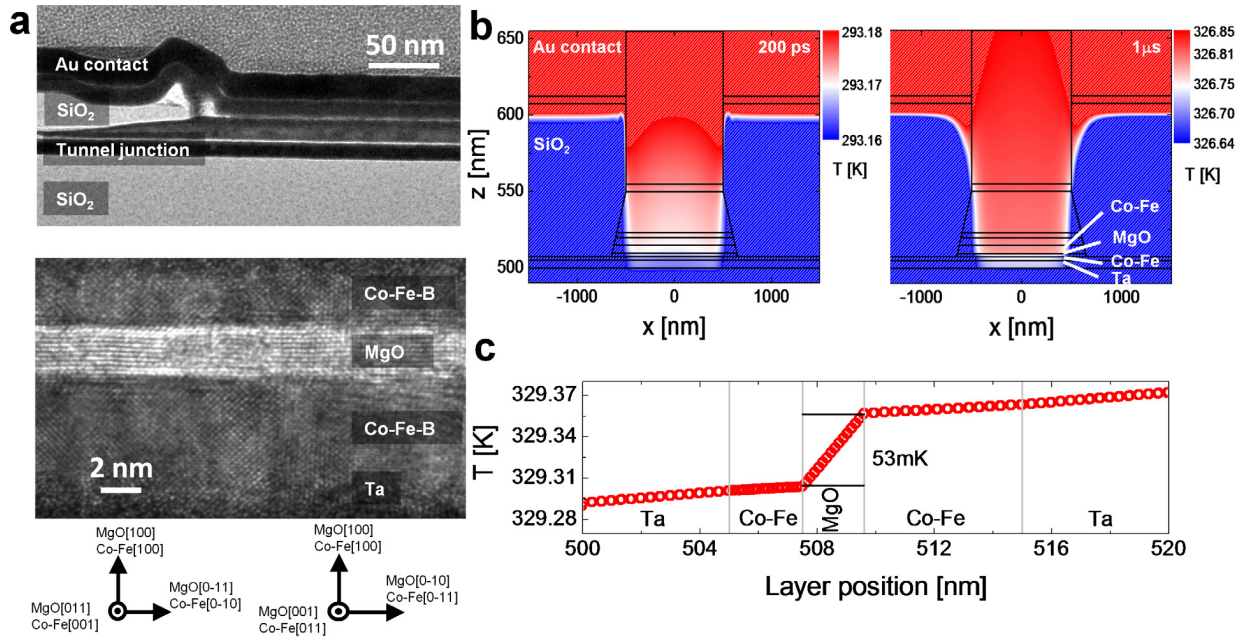

**Figure 3 Cross sections and temperature gradients in the tunnel junction.** a) TMR junctions: device structure studied with Transmission Electron Microscopy (TEM). The high resolution shows the epitaxial relationship Fe-Co(001)/MgO(001) for two transmission directions MgO[100] and MgO[110]. b) On the right side, the simulated temperature distributions are shown for 200 ps and 1 μs after the laser power of 30 mW is turned on in a two dimensional cross section. c) The temperatures for the final static equilibrium condition are shown below as a line scan.

For the experiments, we use Co-Fe-B/ MgO/ Co-Fe-B pseudo spin valve structures. The 1x1 µm$^2$ tunnel junction is heated homogeneously by 30 mW laser power (diode laser with 15-20 µm focus in diameter and a wavelength of 784 nm, Fig 2 a, b) and the charge-Seebeck voltage (Seebeck voltage in the following) is measured for the parallel and antiparallel orientation of the layer magnetization (Fig. 2 c, d). To obtain the temperature distribution and the time constants for the heat diffusion, we used finite element simulations. Transmission Electron Microscopy (TEM) in Fig. 3 a) reveals the device geometry that serves as an input to integrate the heat diffusion equation. In order to calculate the Seebeck coefficients, we estimate a temperature difference ΔT at the 2.1 nm MgO barrier. For polycrystalline MgO films with a nanometer grain size, the heat conductance is lower than the bulk value due to the grain boundaries [15]. The high resolution TEM in Fig. 3a, however, reveals a good crystalline quality of the investigated samples. Nevertheless, the thermal resistance at the Co-Fe/MgO interfaces can have similar effects as the grain boundaries. Therefore, we used both the bulk as well as the reduced value for the thermal conductivities as given in the supplementary information.

In Fig. 3 b) we show the resulting temperature profile in a two dimensional cross section for 200 ps and 1 µs after the laser power is turned on. A series allows determining the time scales of the heating: the static temperature profile is reached after about 2 µs. The final temperature distribution is shown as a line scan in Fig. 3 c) across the tunnel junction. A temperature difference at the 2.1 nm MgO barrier of 53 mK (4.4 mK) is derived from the numerical simulation using the thin film value (bulk value) of the thermal conductivity of MgO respectively.

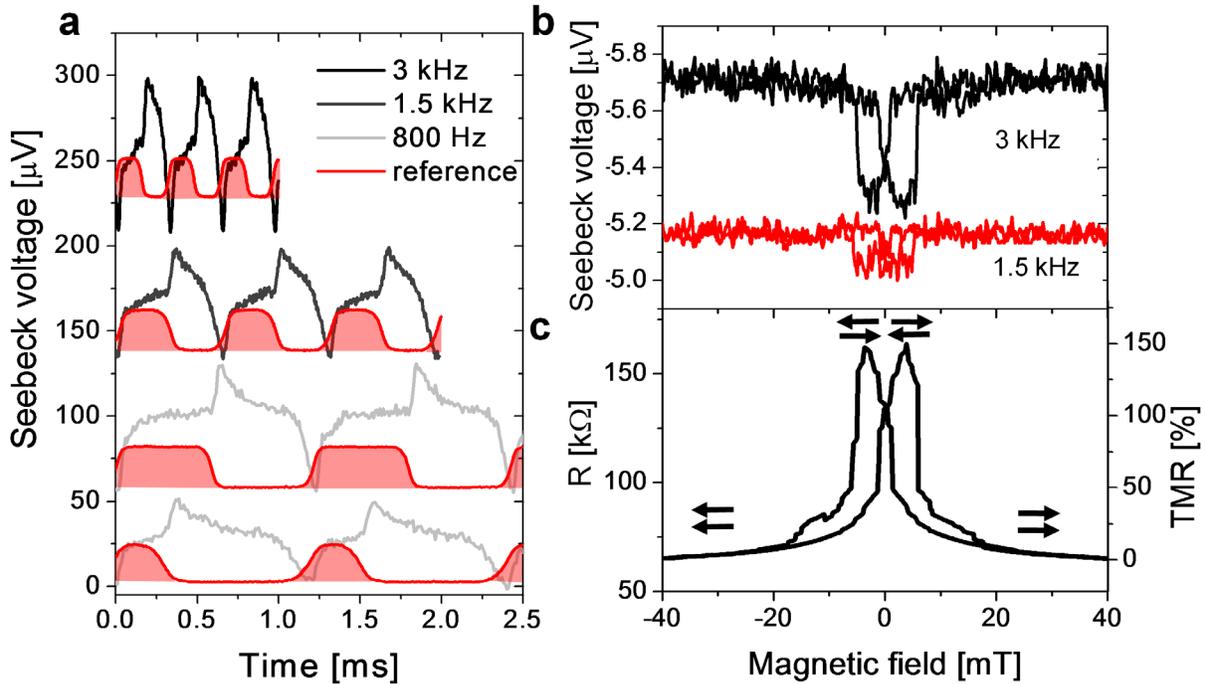

**Figure 4 Seebeck voltages for Fe-Co-B/ MgO/ Fe-Co-B elements.** a) Temporal traces are shown for different modulation frequencies. The heating by the laser is shown in red overlaid to the corresponding Seebeck voltage (the voltages are shifted for clarity). b) Magneto-Seebeck voltages are given above for 1.5 and 3 kHz lock-in modulation. c) The corresponding magnetoresistance (TMR) shows a hard-soft switching of the pseudo spin valve structure.

Fig. 4 shows the magneto-Seebeck effect of a single MTJ with a TMR of 150 %. The temporal voltage traces in Fig. 4 a), as observed in several junctions, show a peak-like voltage when the laser heating is increased and decreased periodically. A negative peak occurs when the laser power is turned on. From the time constants simulated we identify this voltage peak with the Seebeck voltage generated at the junction. The shutter moving through the laser spot limits the time scale to approximately 10-100 µs in our data. As a reference the sequence of the measured laser power is given for each signal trace. The voltage reverses sign when the laser heating is turned off. The Au pads efficiently conduct the heat away from the heat spot (extension of 17.5 µm) into the large bond pads that act as a heat sink. The other side of the junction is then still at a higher temperature and the temperature gradient is reversed. In the lowest curve with a 800 Hz modulation frequency, the magnetic tunnel junction was heated asymmetrically allowing a longer cooling time. The Seebeck voltages for the parallel and antiparallel configurations are determined from Fig. 4 b), in which the Seebeck voltage is shown as a function of the applied field. As expected from the temporal traces, a larger value was found for the fastest modulation of the laser power (at 3 kHz). The Seebeck voltage at the junction contributes more to the total signal than it does for the slower modulation, where the whole sample heats up on a larger area. A signal proportional to the modulation frequency is the dominant component. One obtains -5.7 µV for the parallel and -5.3 µV for the antiparallel orientation, i.e., a change of the Seebeck voltage by 0.4 µV for the tunnel junction. Fluence dependent experiments suggest that the increase of the Seebeck voltage with laser power depends on the increase of the temperature gradient at the barrier and the base temperature at the junction that is increased by the laser power as well, which is discussed further in the supplementary information. If the junction barrier is pushed through a dielectric breakdown [16], the magneto-Seebeck effect disappears.

The experimental results and the theoretical predictions for the Seebeck coefficients are summarized in Table 1. The theoretical prediction for the $Co_{0.5}Fe_{0.5}$ case with Co:Fe 1:1 at the interface is closest to the experiment. The values are negative for both the parallel and the antiparallel configuration. To calculate the Seebeck coefficients from the experimentally determined Seebeck voltage $V_{P,AP}$, we take the temperature gradient to be 53 mK across the 2.1 nm MgO tunnel barrier from our numerical modeling. Thus, one obtains a value of $V_P/\Delta T$= -108 µV/K for the Seebeck coefficient $S_P$ for the parallel orientation.  A decrease of the thermal gradient taking the bulk value for the MgO thermal conductivity as input parameter in our model increases the Seebeck coefficients calculated accordingly. This allows deriving an upper limit of -1300 µV/K for the Seebeck coefficient, given in brackets. Note that spurious other voltages generated in the layer stacks or within the heated device change the Seebeck effect amplitude, but not the difference of the Seebeck voltage for parallel and antiparallel configuration. In accordance with this, the experimental results for the difference ($S_P$-$S_{AP}$) of -8.7 µV/K are closer to the predicted value of -12.1 µV/K than the individual values of $S_P$ and $S_{AP}$. The lower limit of the magneto-Seebeck effect, i.e., the relative change of the Seebeck voltage for the parallel and the antiparallel case, is -8.8 %.

Finally, the magneto-Seebeck effect in magnetic tunnel junctions allows controlling these effects. As a major strategy to develop the possibilities opening up with the magneto-Seebeck effect it is crucial to tailor the thermal tunneling current arising from the majority and the minority spins, i.e. to maximize the shift of the geometric center relative to the Fermi level of these electronic states contributing to the thermal transport. The calculations demonstrate that even the sign of the magneto-Seebeck effect can be controlled using different Co-Fe compositions. Findings on tunnel junctions using a different method (resistive heating) yielding the same effect magnitudes with different sign were recently reported [17] using Co-Fe-B/ MgO/ Co-Fe-B devices however of different composition and structure (using a Singulus Tech. cluster tool). The qualitative change for the devices presented in our letter however allows a comparison of theoretical and experimental results. The results presented compare well to the theoretically predicted change, including the predicted sign reversal of the magneto-Seebeck effect at elevated temperatures (see supplementary information). Further, the experiments showed that the magneto-Seebeck effect can be generated over length scales of only a few nanometers - across a 2.1 nm thick tunnel barrier in our case. This reveals that the magneto-Seebeck effect in magnetic tunnel junctions is a novel effect that can be used to manipulate and design thermovoltages in nanometer-scale devices. The contrast for switching the voltage can be increased further in the future which allows a control of the Seebeck effect by magnetic switching.

**Methods**

Fabrication:

The Co-Fe-B films were prepared by magnetron sputtering using 2-inch targets with compositions of $Co_{0.4}Fe_{0.4}B_{0.2}$ (analysis Co:Fe 0.52:0.48) and $Co_{0.2}Fe_{0.6}B_{0.2}$ (analysis Co:Fe 0.32:0.68) in a UHV system with a base pressure of $5\times10^{-10}$ mbar. They are annealed ex-situ at temperatures of 450°C up to 550°C (post growth annealing, 20-60 min). Samples prepared in the Göttingen chamber were e-beam evaporated MgO after transferring in a separate UHV chamber with base pressure of $5\times10^{-10}$ mbar (maximum TMR reached is 200% at RT). With the Bielefeld chamber, MgO was prepared by magnetron sputtering (maximum TMR reached is 330% at RT). The sample stack was reduced to a simple pseudo spin valve structure to minimize the contribution of spurious Seebeck voltages at metal interfaces not stemming from the junction: Au 27 nm/Ru 3 nm/Ta 5 nm/Co-Fe-B 5.4 nm/MgO 2.1 nm/Co-Fe-B 2.5 nm/Ta 5 nm/$SiO_2$ 500 nm/Si(100). This was done as a trade-off with the magnetic separation of the switching fields, because no anti-ferromagnetic exchange bias layer is used which could also give a magnetic contribution. After ex-situ annealing in a constant field, further structuring was done by standard ion beam etching to yield 1x1 µm$^2$ to 12.5x12.5 µm$^2$ junctions to the MgO barrier. The high resolution transmission electron microscopy (HR-TEM) data in Fig. 4 (bottom, left) reveal the coherent growth of crystallized Co-Fe(110) at each side of the MgO(100) barrier (solid state epitaxy) in columns that can be identified (MgO[001] and MgO[110] in the transmission direction). As an isolation layer at the sides of the element, a 100-nm-thick $SiO_2$ layer was grown by thermal evaporation. A 100 nm thick, top-contact Au layer was deposited as a bond pad. This also prevents direct optical carrier excitation in the MgO barrier. A 5 nm layer of Cr was deposited below the Au top-contact layer for better adhesion on the $SiO_2$ isolation.

Experimental setup:

For the laser heating, a 100 mW, Toptica, intensity-stabilized laser diode module ($\lambda$ = 784 nm) was focused to a diameter of 15-20 µm full width at half maximum (FWHM). For the standard experiments we used 30 mW laser power. The beam position on top of the bond pad was controlled via a camera. The intensity was modulated using 800 Hz, 1.5 kHz, and 3 kHz modulation frequency. To prevent a current flow in the system that could be modified by the change of the resistance of the junction, a high input impedance (100 GΩ) LT1113 precision operational amplifier (Linear Technology) was used. The band width of the amplifier is 5 MHz. This was installed close to the sample to minimize the effect of the cable capacitance (< 15 pF). A simulation of the circuit with the sample resistance showed that the change of the resistance will contribute < 1 nV to the absolute voltage. Seebeck voltage versus magnetic field curves is measured using a Stanford Linear research lock-in amplifier.

Thermal modeling of parameters (COMSOL):

To simulate heat flow and temperature distribution, the MTJ was modeled using the COMSOL finite-element package. The tunnel junction geometry was taken from the cross sectional TEM data as input parameters. The element was embedded into a 3 µm cylinder. The heat flow from the laser heating comes from the top. For a 30 mW laser power the absorbed laser power is 10 mW. The temperature at the bottom of the cylinder in the Si(100) substrate was set to ambient temperature. The 500 nm $SiO_2$ layer on top of the substrate is the bottleneck for heat diffusion through the cylinder stack. The temperature at the bottom layer of the element depends sensitively on the heat flow though $SiO_2$ layer and determines the 2 µs needed to reach the final heat gradient. A prism-shaped, adaptive mesh was used with about >10 nm in the plane and sub-nanometer resolution perpendicular to the plane. In addition, we performed simulations on the larger length scale to simulate lateral heat diffusion. The heated area extends to about a 17.5 µm diameter, and, in that case, the absolute temperature increase is reduced to 8 K. The equilibrium heat gradient is attained within about 2 µs. For the 2.1 nm MgO thin tunnel barrier, a value of the thermal conductance of $\kappa$ = 4 W/(m·K) is assumed to be closest to the reality. This value has been determined experimentally for a thin film [15]. It is expected to be much closer to the bulk value of $\kappa$ = 48 W/(m·K) which gives an upper limit for the Seebeck coefficient. All material parameters used in the numerical model are provided in a table in the supplementary information.

**Author contributions**

**Acknowledgments**

A.T. acknowledges the Ministry of Innovation, Science and Research (MIWF) of the North Rhine-Westphalia state government for financial support. M.M., M.S., M.W., and P.P. acknowledge the funding provided by the German research foundation (DFG) via the SFB 602 for the TEM work. M.C., M.B., and C.H. acknowledge support from DFG SPP 1386 and DFG grant HE 5922/1-1. J.S.M acknowledges support by the US National Science Foundation (NSF) and Office of Naval Research (ONR). We acknowledge Anissa Zeghuzi's help for additional COMSOL calculations. This work was initiated by the SpinCaT priority program.

# Seebeck Effect in Magnetic Tunnel Junctions

**Supplementary information**

1. Details of the laser heating and Seebeck voltage transients

The slowest time constant in the heating process arises from the low conducting electric isolation SiO$_2$ layer on the Si wafer, which is the thermal bottleneck. The temperature gradient at the barrier itself, which is responsible for the voltage generation, is 4.4 mK, in accordance with the simple estimation using the ideal bulk value for MgO for the thermal conductivity $\kappa$. A closer fit of theoretical and experimental results will be possible with a more precise experimental determination of the temperature gradient at the MgO barrier, since it depends sensitively on the thermal conductivity of the thin films and interfaces and thus has the largest error bar. While, in principle, this task is possible for such thin films [1], it will be difficult to accomplish especially in a single device. However the thin film value for the 2.1 nm MgO layer is expected to be less than its bulk value of $\kappa$ = 48 W/(m·K). For polycrystalline MgO films with a grain size of 3-7 nm, a value of $\kappa$ = 4 W/(m·K) was found at room temperature, which differs by a factor of ten [2]. This difference arises from the grain boundaries. In addition, the thermal resistance at the Co-Fe/MgO interface can be considerable, i.e., 1x10$^9$ W/(m$^2$·K) [1]. This interface term is approximately of the same order as the thermal resistance of a film that is 1 nm thick and may further increase the thermal gradient and thus decrease the Seebeck coefficients calculated. Taking into account the crystalline quality of the tunnel junctions grown by solid-state epitaxy as derived from the high resolution TEM, the most realistic scenario is a model assuming bulk-like thermal conductivity for the MgO because of its crystalline quality and a thermal interface resistance at both Co-Fe/MgO interfaces. Thus the value for the MgO tunnel barrier is expected to be between the thin film value for the polycrystalline material and the bulk value. Here we give the polycrystalline and the bulk value as a lower (conservative estimation) and upper limit (optimistic estimation).

| Material | $\kappa$ [W/(m·K)] | $\rho$ [10$^3$ kg/m$^3$] | $c_v/c_p$ [J/(kg·K)] |
|---|---|---|---|
| Ta | 57 | 16.7 | 140 |
| Ru | 117 | 12.4 | 238 |
| Au | 320 | 19.3 | 128 |
| Cr | 94 | 7.2 | 449 |
| MgO | 4 (48) | 3.6 | 935 |
| SiO$_2$ | 1.4 | 2.2 | 1052 |
| Co$_{20}$Fe$_{60}$B$_{20}$ | 87 | 8.2 | 440 |

Table 1: For MgO the thin film value (bulk value) of the thermal conductivity given respectively.

In the following, we discuss the dynamic time scales that contribute to the voltage response to the laser heating experiments shown for different modulation frequencies in Fig. 4a (main manuscript). Time traces have been measured after amplification (100-MHz PM 3382 Phillips oscilloscope). A negative voltage was found for the opening of the shutter upon heating. When the shutter is closed, the cooling reverses the sign of the voltage. The heat is effectively conducted away from the 17.5 μm heat spot into the large 100 nm thick Au bond pads. The time constant of the voltage increases within a few μs, i.e., the response of the voltages are approximately 10-100 µs, generally asymmetric for the heating and cooling process. The dynamic response depends on three factors, all of which were in the microsecond regime, i.e., 1) the time constant of heating, 2) the width of the Gaussian laser intensity profile that moves across the chopper blade, and 3) the response time of the LT1113 precision operational amplifier to realize a high input impedance of 100 GΩ. While the thermal equilibration time in the COMSOL model is 2 µs, and the response time of the amplifier 5 MHz, we identified the width of the Gaussian laser intensity profile that moves across the chopper blade as slowest contribution limiting the response time. The laser diameter and chopper blade dimensions (different for each modulation frequency with varied chopper blades) range from 20-150 µs and can be extracted from the reference diode signal in Fig. 4a (main manuscript).

In addition, we conducted tests to determine whether spurious magneto voltages were generated in the layer stack. A natural failure of the barrier (dielectric breakdown) at low voltages allows shorten the junction without too much destruction of the element and the magneto-Seebeck effect should vanish. To prove this claim, we conducted tests to determine whether magneto-Seebeck voltage was generated at the tunnel barrier. The tests consisted of gradually shorting the barrier, as evidenced by a decrease in the resistance from the 100 kΩ of an intact junction to 1 kΩ and, finally, to 100 Ω for the shortened junction. In the intermediate state, the magneto-Seebeck effect decreased to about 1%, but the magnetic effect in the Seebeck voltage vanishes completely if the MgO tunnel barrier is bridged. This excludes the existence of magneto voltages not generated at the tunnel barrier. However, a background voltage is still contributing to the signal, thereby limiting the experimentally determined value of the magneto-Seebeck effect, $S_{MS}$. Also, we can exclude the generation of lateral spin currents. This was checked by reversing the direction of rotation of the chopper blade and thereby changing the direction of the possible lateral heat gradients in plane. This sign change should in that case also lead to sign reversal which is not observed.

[1]  David G. Cahill, Analysis of heat flow in layered structures for time-domain thermoreflectance. Rev. Sci. Instrum. **75**, 5119 (2004).
[2]  S.M. Lee and D.G. Cahill, T.H. Allen, Thermal conductivity of sputtered oxide films. Phys. Rev. B 52, 253-257 (1995).

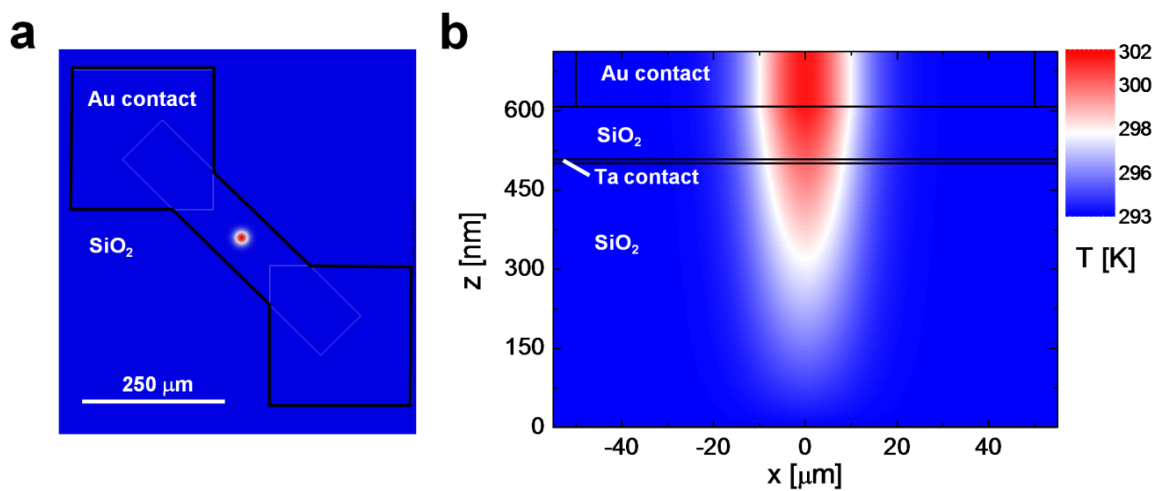

Fig. 1: COMSOL simulation on the larger length scale to illustrate the extension of the heating spot of 15 µm. The diameter of the spot heated up extends to 17.5 µm.

## 2. Lateral heating in the Au bond pad

Shown in Fig. 1 is the full bond pad that has been simulated to reveal the extension of the heated area. The simulation shows that the Au bond pad is not heated homogeneously. The Gaussian beam diameter of 15 µm results in a heated area of 17.5 µm in diameter, not much further extended than the area heated by the laser spot. The reason is the large heat conductivity of 320 W/(m·K) of the 100 nm thick Au bond pad layer. During one modulation cycle first, the laser power is turned on the temperature rises. Second, when turned off, the large Au bond pads that have the highest heat conductivity work as a cooler. The temperature gradient is reversed in the junction. This is detected in the sign reversal of the Seebeck voltage for the transient voltage traces.

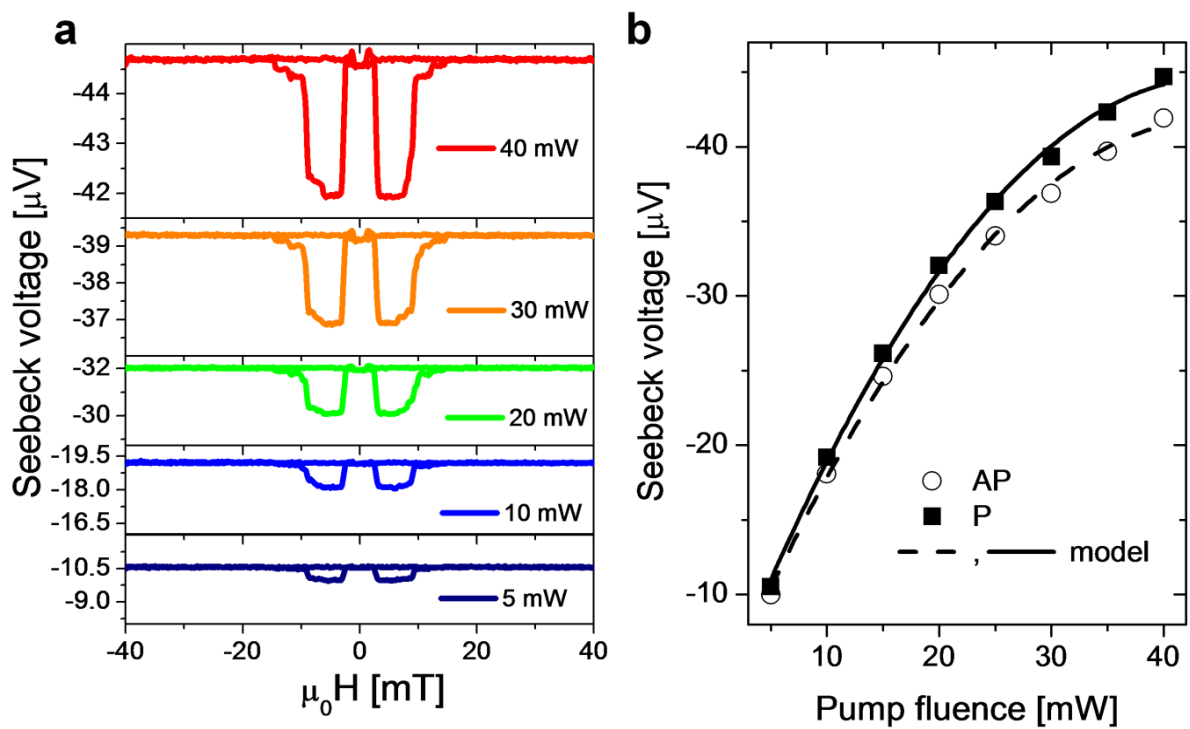

Fig. 2: Seebeck voltage measured for a tunnel junction heated with low laser fluencies (diode laser). a) Seebeck voltage versus applied magnetic field. b) Seebeck voltages for parallel and antiparallel orientation. The model is derived assuming a change of the Seebeck coefficient with temperature.

## 3. Fluence dependence of the Seebeck voltages

In the following the fluence dependence of the Seebeck voltage is discussed. The magnetoresistance of the junction is 120 % and it shows a magneto-Seebeck effect of 6.2 %, which is somewhat smaller than for the element discussed in the main manuscript in particular. The voltage change from parallel to antiparallel configuration is 2.5 µV (for 30 mW laser power). In the experiment the laser fluence is varied from 5 to 40 mW. It can be seen that with increasing the laser fluence the shape of the signal does not change. However the Seebeck voltage increases as expected. For the highest laser fluence one finds -45 µV in the parallel configuration and -42 µV for the antiparallel configuration. To discuss the evolution in detail, the values are given in Fig. 2. One observes that the voltage increases nonlinearly with the pump fluence. However this behavior is expected: with increasing the pump-fluence not only the temperature difference at the barrier increases, also the base temperature increases. One can calculate the voltage dependence from the following model

$$V_{P,AP} = S_{P,AP}\Delta T = s_{P,AP} T\, a\,(T - 293K) \quad (1)$$

Here $s_{P,AP}$ is defined as the slope of the $S_{P,AP}$ versus T curve. The parameter $a$ is given by the total junction geometry. It relates the temperature gradient at the junction ($\Delta T$) to the temperature gradient in between the tunnel barrier (T) and room temperature (293 K) at the substrate bottom. From a series of simulations using laser fluences from 5 to 40 mW, we estimate a value of 1.8 mK/ mW as rise in the temperature gradient ΔT with laser power. The increase of the temperature at the tunnel barrier is 1.2 K/ mW with laser power. In addition, the theoretical model predicts that the Seebeck coefficient itself decreases as the base temperature is raised. Assuming a constant slope $s_{P,AP}$, we can qualitatively understand the linear plus quadratic behavior of the data in Fig. 2b.

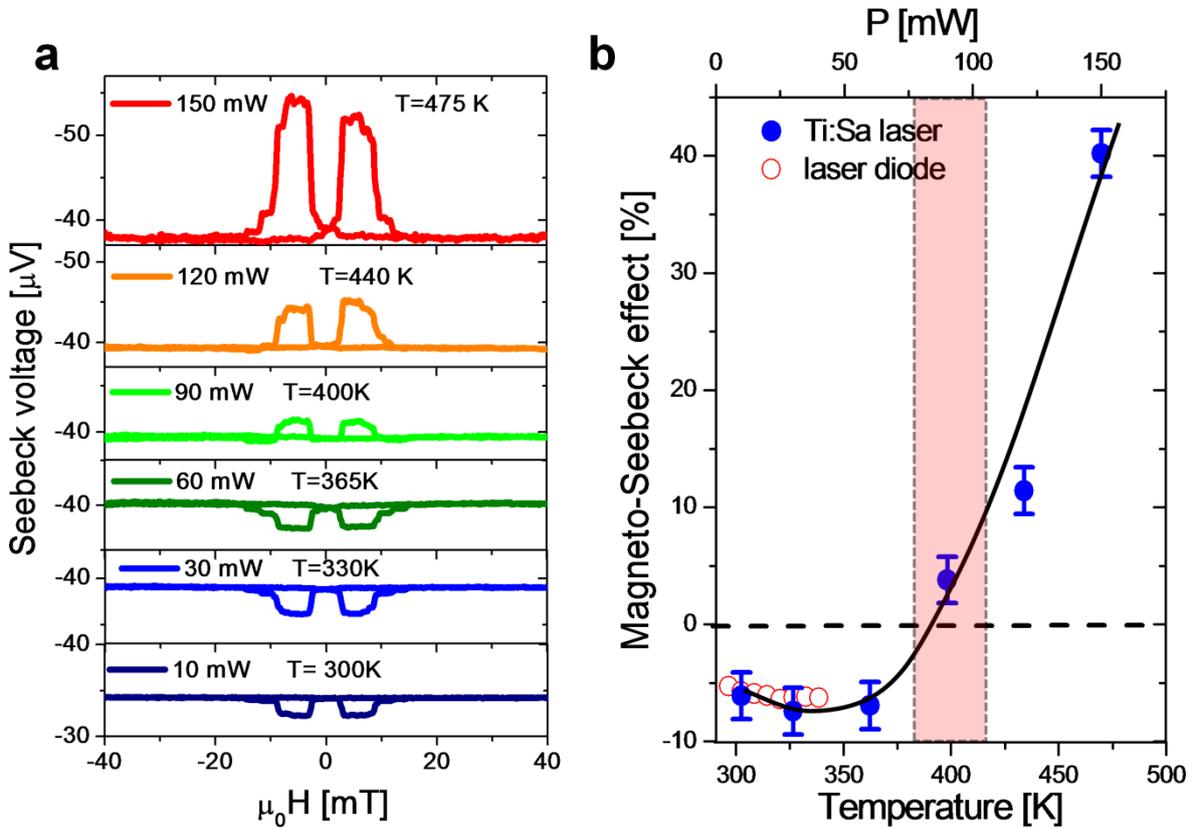

Fig. 3: Seebeck voltage measured for a tunnel junction heated with high laser fluences (Ti:sapphire oscillator, central wave length λ=810 nm). a) Seebeck voltage versus applied magnetic field. b) Magneto-Seebeck effect as a function of the laser power and corresponding temperature calculated by the numerical model. The data for the diode laser is shown for direct comparison. A sign change is observed in the region at around 400 K as expected from the theoretical model.

Even larger fluences are reached using a Ti:sapphire laser, which can be regarded here as a cw laser source at moment. This is justified to a good approximation because of the large distance from the Au top layer to the tunnel barrier (>130nm). We find that the magneto-

Seebeck values for high laser fluences lie on top of the previous data determined using the cw diode laser, both shown in the shown in the same figure for direct comparison (Fig. 3b). One observes a characteristic sign change of the magneto-Seebeck effect in-between 60 and 90 mW. This sign change is predicted our theoretical model (Fig. 1 e, main manuscript). By relating the pump fluence to a temperature using the numerical finite element simulation, one can plot the magneto-Seebeck effect as a function of the temperature derived at the tunnel junction (Fig. 3b). For the largest laser power the base temperature can be increased up to 475 K. The sign reversal takes place at around the predicted compensation point for the Seebeck voltages for parallel and antiparallel orientation at 400 K. However there are also features observed not expected from the model. The magneto-Seebeck effect shows a decrease in between 300 to 350 K before it rises and crosses zero. However, the magneto-Seebeck effect reaches 40% for 150 mW laser power, close to the theoretically expected value. The maximum voltage change from parallel to antiparallel configuration is 15 $\mu$V. Altogether the finding of the prominent feature, the sign reversal with increased temperature, supports the conclusion that the transport coefficients calculated according to the model of coherent tunneling model can explain characteristic features and thus is expected to be the dominating contribution to the magneto-Seebeck effect.

In the high fluence range one may ask if for such large temperature gradients the theoretical model is still valid. The temperature gradient at the 2.1 nm thick MgO barrier is 270 mK for the highest fluence. To answer this question, it is helpful to compare this temperature to an energy scale. Bias dependence is calculated in such system for up to 1 V bias voltage [3]. This amounts to 1 eV disturbance of the electrochemical potential or huge voltage gradients of 10.000.000 V/cm. Therefore the temperature gradients here can be regarded as a much weaker disturbance.